\documentclass[12pt]{article}
\usepackage{arxiv}

\usepackage{textcomp} 
\usepackage[utf8]{inputenc} 
\usepackage[T1]{fontenc}    
\usepackage{hyperref}       
\usepackage{url}            
\usepackage{booktabs, threeparttable}       
\usepackage{amsfonts, amssymb, amsmath, latexsym, amsthm, bm}       
\usepackage{nicefrac}       
\usepackage{microtype}      
\usepackage{lipsum}		
\usepackage{graphicx}
\usepackage{natbib}
\usepackage{doi}
\usepackage{pdflscape}
\usepackage{listings}
\title{surveygenmod2: A SAS macro for estimating complex survey adjusted generalized linear models and Wald-type tests}
\author{ 
        {R. Noah Padgett} \\
	Department of Epidemiology\\
	Harvard T.H. Chan School of Public Health\\
        Harvard University\\
	Boston, MA  \\
	\texttt{npadgett@hsph.harvard.edu} \\
	\And
	{Ying Chen} \\
	Department of Epidemiology\\
	Harvard T.H. Chan School of Public Health\\
        Harvard University\\
	Boston, MA \\
	\texttt{yingchen@fas.harvard.edu} \\
}

\renewcommand{\shorttitle}{surveygenmod2}

\hypersetup{
pdftitle={surveygenmod2: A SAS macro for estimating complex survey adjusted generalized linear models and Wald-type tests},
pdfsubject={stat.ME},
pdfauthor={R. Noah Padgett, Ying Chen},
pdfkeywords={survey methods, taylor series linearization, global flourishing study, SAS, surveygenmod},
}

\begin{document}
\large
\maketitle

\begin{abstract}
surveygenmod2 builds on the macro written by da Silva (2017) for generalized linear models under complex survey designs. The updated macro fixed several minor bugs we encountered while updating the macro for use in SAS\textregistered. We added additional features for conducting basic Wald-type tests on groups of parameters based on the estimated regression coefficients and parameter variance-covariance matrix.
\end{abstract}

\keywords{survey methods \and taylor series linearization \and global flourishing study \and SAS \and surveygenmod \and modified Poisson \and Wald tests}

\section*{Introduction}

Data gathered in modern survey typically fall under the heading of a ``complex survey'' because the design of the survey is set up to reduce the burden on sampling and ensure coverage of the population \citep{Groves2004}.
Failing to account for the characteristics of the sampling design will lead to biased estimation and underestimation of the uncertainty in your estimates \citep{Little2008}.
Therefore, utilizing methods that incorporate the complex sampling design into the analysis is crucial to obtain valid inferences.

The use of generalized linear models is fundamental to the modeling and analysis discrete data, such as binary or count outcomes \citep{Agresti2012}.
And incorporating the complex sampling design into the modeling of these types of outcomes can be challenging because the estimators rarely have simple forms and are commonly nonlinear leading to computational complexities \citep{Wolter2007}.

The current work builds on the macro developed by \citet{daSilva2017} for estimating generalized linear models under complex survey designs in SAS\textregistered.
The remainder of the whitepaper is organized as follows.
Next, we summarize the technical details of underlying the estimation.
Then, we provide an simple example of a complex survey adjusted modified Poisson regression.

\section*{Generalized Linear Models under Complex Sampling}

Generalized linear models (GLM) unify many seemingly distinct types of models into a umbrella with common features, namely, (1) a distribution which governs the randomness of errors, (2) systemic differences modeled by a design matrix that is linear (e.g., in the form $\mathbf{X} \bm\beta$), and (3) a link function ($g(.)$) which joins the random and systematic parts together. Together, a simple representation of the GLM is:
$$\mathbf{y}=g\left(\mathbf{X}\bm\beta\right)+\bm\epsilon.$$
The link function needs to be a relatively simple location-scale transformation defined by a monotonic function.
The transformed linear combination of the design matrix and regression coefficients is commonly denoted by the simplified notation $\bm\eta=g\left(\mathbf{X}\bm\beta\right)$.

\subsection*{Estimation}
Following the discussion from \citet{daSilva2017}, the primary method for parameter estimation is pseudo-maximum likelihood. 
Pseudo-maximum likelihood is regular maximum likelihood with the likelihood function is weighted by the sampling weight associated with each case.
The weighted log-likelihood for an arbitrary exponential distribution response function ($f(.)$) is:
\begin{equation}
    l(y,\mathbf{X}, \bm\beta, \phi, \mathbf{w}) = \sum_{\forall i} \log\left(f(y_i ; \mathbf{x}_i,\bm\beta, \phi, w_i\right),
\end{equation}
where $w_i\in \mathbf{w}$ is the vector of sampling weights. When all weights are equal to 1, the standard maximun likelihood estimators are obtain. Incorporating the sampling weights ensures the point estimates are unbiased--provided the weights are property calibrated to the population.
For estimation, a ridge-stabilized Newton-Raphson algorithm is used which is the same method employed by \texttt{proc genmod} \cite{sas2011}.
In the k-th iteration, the vector of parameter estimates ($\bm\beta_k$) is updated using the inverse Hessian ($\mathbf{H}^{-1}$) and gradient of the pseudo log-likelihood ($\mathbf{s}$), that is:
\begin{align}
    \bm\beta_{k+1} &=\bm\beta_{k} - \mathbf{H}^{-1}\mathbf{s}\\
    \mathbf{H} &= \left[h_{ij}\right] = \left[\frac{\partial^2L}{\partial\beta_i\partial\beta_j}\right]\\
    \mathbf{s} &= \left[s_j\right] = \left[\frac{\partial L}{\partial \beta_j}\right].
\end{align}

For models with some scale and/or dispersion parameter such as normal, gamma, negative binomial, and inverse normal, these parameters are assumed to exist and estimated via maximum likelihood or by method of moments \cite{daSilva2017}. To estimate the vector of gradients ($\mathbf{s}$) and hessian matrix ($\mathbf{H}$), we can use the chain rule of differentiation of obtain the elements of each matrix/vector index using the corresponding cells of the matrix/vector. Simplifying the notation by using $\mu_i=g^{-1}(\mathbf{x}_i^t\bm\beta)$, the gradient and hessian are
\begin{align*}
    \mathbf{s} &= \sum_{\forall i} \frac{w_i(y_i-\mu_i)}{V(\mu_i)g^\prime(\mu_i)\phi} \mathbf{X}^t,\\
    \mathbf{H} &= -\mathbf{X}^t\mathbf{W_0}\mathbf{X},
\end{align*}
where $V(.)$ is the variance function; $\phi$ is the distribution specific parameter of dispersion; and $\mathbf{W_0}$ is a diagonal weight matrix. The elements of the weight matrix are
\begin{align*}
    \mathbf{W}_0[i,i] &= w_{ei} + w_i(y_i-\mu_i)\frac{V(\mu_i)g^{\prime\prime}(\mu_i)+V^{\prime}(\mu_i)g^\prime(\mu_i)}{\left(V(\mu_i)\right)^2\left(g^\prime(\mu_i)\right)^3\phi}\\
    w_{ei} &= \frac{w_i}{V(\mu_i)\left(g^\prime(\mu_i)\right)^2\phi}.
\end{align*}
The information matrix is given by the negative of the observed hessian. The sampling weight $w_i$ can be interpreted as the prior weight of each case, but the case weights are then updated during estimation of the parameters to ensure the cases with less precision are down weighted in the estimation of point estimates to obtain variance stabilized estimates of parameters.

\subsection*{Variance Estimate}

The variance estimation similarly followed from da Silva \cite{daSilva2017}, where the parameter variance-covariance matrix is estimated using the Taylor series linearization approach to help obtain more consistent variance estimators. The estimated parameter variances are
\begin{align*}
    \hat{\mathbf{V}}\left(\hat{\bm\beta}\right) &= \hat{\mathbf{Q}}^{-1}\mathbf{G}\hat{\mathbf{Q}}^{-1}\\
    \hat{\mathbf{Q}} &= \phi\mathbf{X}^t\mathbf{W}_e\mathbf{X} = \sum_{h=1}^{H}\sum_{i=1}^{n_h}\sum_{j=1}^{m_{hi}} w_{hij}\hat{\mathbf{D}}_{hij} \left(V(\mu_{hij})\right)^{-1}\hat{\mathbf{D}}^t_{hij}\\
    \hat{\mathbf{G}} &= \frac{n-1}{n-p}\sum_{h=1}^{H}\frac{n_h(1-f_h)}{n_h-1}\sum_{i=1}^{n_h}{(e_{hi.}-\Bar{e}_{h..})}{(e_{hi.}-\Bar{e}_{h..})}^t\\
    e_{hi.} &= \phi\mathbf{W}^{\ast}_e(\mathbf{y}-\bm\mu)\mathbf{X} = \sum_{j=1}^{m_{hi}}w_{hij}\hat{\mathbf{D}}_{hij} {\left(V(\mu_{hij})\right)}^{-1}(y_{hij}-\hat\mu_{hij})\\
    \Bar{e}_{h..} &=  \frac{1}{n_h}\sum_{i=1}^{n_h}e_{hi.}\\
    \hat{\mathbf{D}}_{hij} &= \left[\frac{1}{g^{\prime}(\mu_{hij})}\right]^{-1}\\
    \mathbf{W}^{\ast}_e &= \text{diag}\left(\frac{w_i}{V(\mu_i)\left(g^\prime(\mu_i)\right)^2\phi}\right).
\end{align*}

\subsection*{Joint Tests for Groups of Regression Coefficients}

A Wald-type test was implemented to help test groups of parameter simultaneous.
These tests used the estimated parameter variance-covariance matrix ($\mathbf{V}$) to test the hypothesis $H_0:\ \mathbf{L}\bm\beta=0$ where $\bm\beta$ is a vector of regression coefficients, and $\mathbf{L}$ is a design-like matrix specifying which elements in $\bm\beta$ are being tested. 
The test statistic computed is
\begin{equation*}
F_{Wald}= \frac{\left(\mathbf{L}\bm{\hat{\beta}}\right)^t\left(\mathbf{L}^t\hat{\mathbf{V}}\mathbf{L}\right)^{-1}\left(\mathbf{L}\bm{\hat{\beta}}\right) }{\text{rank}\left(\mathbf{L}^t\hat{\mathbf{V}}\mathbf{L}\right)}.    
\end{equation*}

Then, $p$-values are obtained from the F-distribution with numerator degrees of freedom equal to the number of tested parameters (or rank of $\mathbf{L}^t\hat{\mathbf{V}}\mathbf{L}$), and denominator degrees of freedom equal to the model degrees of freedom from the regression analysis minus the number of parameters tested (i.e., the sum of weights to approximate sample size minus the rank of $\mathbf{L}^t\hat{\mathbf{V}}\mathbf{L}$). 

\section*{Example Analyses}

\subsection*{Global Flourishing Study Data}

The description of the methods below have been adapted from \cite{VanderWeele2024}. 
Currently available data from Wave 1 of the Global Flourishing Study includes nationally representative samples of the adult population from 22 geographically and culturally diverse countries, including Argentina, Australia, Brazil, Egypt, Germany, Hong Kong (Special Administrative Region of China), India, Indonesia, Israel, Japan, Kenya, Mexico, Nigeria, the Philippines, Poland, South Africa, Spain, Sweden, Tanzania, Türkiye, United Kingdom, and the United States (Wave 1 data will also be available for mainland China once Wave 2 data are released in early 2025). These countries were selected to (a) maximize coverage of the world's population, (b) ensure geographic, cultural, and religious diversity, and (c) prioritize feasibility and existing data collection infrastructure. Data collection was carried out by Gallup, a global analytics and advisory organization with decades of experience collecting global data on various aspects of human life. Most of the data for Wave 1 were collected in 2023, with some countries beginning data collection in 2022; exact dates of data collection vary by country \cite{Ritter2024}. The GFS is set to continue with four additional waves of annual data collection from 2024-2027. The precise sampling design varied by country to ensure nationally representative samples for each country. Further details of the sampling design methodology are available elsewhere \citep{Padgett2024, Ritter2024}.

Survey items included numerous aspects of well-being such as happiness and life satisfaction, physical and mental health, meaning and purpose, character and virtue, close social relationships, and financial and material stability \cite{VanderWeele2017}, along with numerous other demographic, social, economic, political, religious, personality, childhood, community, health, and well-being variables. The data are publicly available through the Center for Open Science (https://www.cos.io/gfs). During the translation process, Gallup adhered to TRAPD model (translation, review, adjudication, pretesting, and documentation) for cross-cultural survey research \cite{Harkness2016}. 

\subsection*{Illustrative Results}

Two example analyses were conducted using the surveygenmod2 macro described in this paper. 
For these examples, we used two different outcomes from the GFS, feelings about expenses ("How often do you worry about being able to meet normal monthly living expenses?") and feelings about income ("Which one of these phrases comes closest to your own feelings about your household's income these days?"). 
A 10\% stratified random sample from the dataset using the Gallup provided strata and weights.
The first analysis is a survey weighted linear regression analysis with normally distributed errors.
The second is a modified Poisson regression analysis with a binary outcome.
A modified Poission regression is a popular approach to estimating risk-ratios in epidemiological work \citep{Zou2004}.
Results are compared with the \texttt{survey} package \citep{survey2024, Lumley2004} in R \citep{R2024}, and \texttt{proc surveyreg} in SAS \citep{SASprocsurveyreg}.

\subsubsection*{Linear Regression}
The approximately continuous outcome of worry/feelings about expenses was analyzed first on the entire GFS dataset using the survey package in R with all adjustments to get a "population" estimate from the GFS data.
\begin{table}[!htp]
    \centering
    \caption{Complex survey adjusted linear regression of expenses across R and SAS}
    \label{tb:linear-reg}
    \begin{tabular}{llrrrrrr}
    \toprule
     & & \multicolumn{2}{c}{SAS: proc surveyreg} & \multicolumn{2}{c}{SAS: surveygenmod2} & \multicolumn{2}{c}{R: survey}\\ \cmidrule(lr){3-4} \cmidrule(lr){5-6} \cmidrule(lr){7-8}
     \multicolumn{2}{l}{Term} &   Est & SE &  Est & SE & Est & SE \\ \midrule
     \multicolumn{2}{l}{(Intercept)} & 5.501 & 0.093 & 5.501 & 0.093 & 5.501 & 0.093\\
     \multicolumn{2}{l}{Age (mean centered)} & 0.016 & 0.002 & 0.016 & 0.002 & 0.016 & 0.002\\
     \multicolumn{2}{l}{Gender (Ref: Male)} &  &  & &  & & \\
      & Female & $-$0.434 & 0.078 & $-$0.434 & 0.078 & $-$0.434 & 0.078\\
      & Other & $-$0.616 & 0.463 & $-$0.616 & 0.463 & $-$0.616 & 0.463\\
     \multicolumn{2}{l}{Education (Ref: 9-15 years)} &  &   &  &    &  & \\
      & 16+ years & 0.859 & 0.082 & 0.859 & 0.083 & 0.859 & 0.083\\
      & Under 8 years & $-$0.681 & 0.089 & $-$0.681 & 0.090  & $-$0.681 & 0.090 \\
     \multicolumn{2}{l}{Self-Rated Health (Ref: Good)} &  &   & &    & & \\
     & Excellent & 0.749 & 0.104 & 0.750 & 0.104  & 0.749 & 0.104 \\
     & Very good & 0.413 & 0.105 & 0.413 & 0.105 & 0.413 & 0.105\\
     & Fair & $-$0.176 & 0.171 & $-$0.176 & 0.171 & $-$0.176 & 0.171\\
     & Poor & $-$1.199 & 0.294 & $-$1.199 & 0.294 & $-$1.199 & 0.294\\ \bottomrule
    \end{tabular}
\end{table}

\subsubsection*{Modified Poisson Regression}
For analyzing feelings about income as a binary outcome, the categories were collapsed as 1 = "Living comfortably on present income/Getting by on present income" and 0 = "Finding it difficult on present income/Finding it very difficult on present income."

\begin{table}[!htp]
    \centering
    \caption{Complex survey adjusted modified Poisson income feelings across R and SAS}
    \label{tb:mod-poisson}
    \begin{tabular}{llrrrr}
    \toprule
     & & \multicolumn{2}{c}{SAS: surveygenmod2} & \multicolumn{2}{c}{R: survey}\\ \cmidrule(lr){3-4} \cmidrule(lr){5-6}
     \multicolumn{2}{l}{Term} &  Est & SE & Est & SE \\ \midrule
     \multicolumn{2}{l}{(Intercept)} & $-$0.799 & 0.036 & $-$0.799 & 0.036\\
     \multicolumn{2}{l}{Age (mean centered)} & $-$0.001 & 0.001 & $-$0.001 & 0.001\\
     \multicolumn{2}{l}{Gender (Ref: Male)} & &  & & \\
      & Female & $-$0.095 & 0.028 & $-$0.095 & 0.028\\
      & Other & 0.060 & 0.250 & 0.060 & 0.250\\
     \multicolumn{2}{l}{Education (Ref: 9-15 years)} &  &   &  &   \\
      & 16+ years & $-$0.017 & 0.030 & $-$0.017 & 0.029\\
      & Under 8 years & $-$0.250 & 0.035  & $-$0.245 & 0.034 \\
     \multicolumn{2}{l}{Self-Rated Health (Ref: Good)} &  &   & &  \\
     & Excellent & 0.070 & 0.039  & 0.069 & 0.039 \\
     & Very good & 0.122 & 0.039 & 0.122 & 0.039\\
     & Fair & 0.034 & 0.061 & 0.034 & 0.061\\
     & Poor & $-$0.413 & 0.123 & $-$0.413 & 0.123\\ \bottomrule
    \end{tabular}
\end{table}

\section*{Concluding Remarks}

The updated surveygenmod2 macro we modified based on the work of da Silva \citep{daSilva2017}.
When referencing this work, please also cite da Silva's original work published as a proceedings of the SAS Global Forum. 
he macro was revised and updated by R. Noah Padgett and Ying Chen of the Human Flourishing Program at Harvard University as part of their contributions to the Global Flourishing Study \citep{GFS2024}.

To access the macro in an easily copied form, please visit our GitHub repository:

\url{https://github.com/noah-padgett/surveygenmod2}

Please add an issue to the repository for feature requests and bugs you find. You can also email the lead contributor, R. Noah Padgett, with any comments and suggestions. 

\subsection*{Commitment to open science practices}
Data for Wave 1 of the GFS is available through the Center for Open Science upon submission of a pre-registration and will be openly available without pre-registration beginning February 2025. Subsequent waves of the GFS will similarly be made available. Please see \url{https://www.cos.io/gfs-access-data}  for more information about data access.
\subsection*{Acknowledgments}

This study is possible thanks to the generous support of: The John Templeton Foundation; Templeton Religion Trust; Templeton World Charity Foundation; The Well-Being for Planet Earth Foundation; The Fetzer Institute; Well Being Trust; The Paul L. Foster Family Foundation; and The David \& Carol Myers Foundation.

\bibliographystyle{apalike}
\bibliography{surveygenmod2}

\end{document}